\documentclass[twocolumn,aps,prb,superscriptaddress,floatfix]{revtex4-2}
\usepackage{amsmath}
\usepackage{amssymb}
\usepackage{graphicx}
\usepackage{color}
\usepackage{hyperref}
\usepackage{bm}

\begin{document}

\title{Geometry-independent tight-binding method for massless Dirac fermions in two dimensions}
 
\author{Alexander Ziesen}\email{alexander.ziesen@rwth-aachen.de} 
\affiliation{JARA Institute for Quantum Information, RWTH Aachen University}

\author{Ion Cosma Fulga} 
\affiliation{IFW Dresden and Würzburg-Dresden Cluster of Excellence ct.qmat, Helmholtzstr. 20, 01069 Dresden, Germany}

\author{Fabian Hassler} 
\affiliation{JARA Institute for Quantum Information, RWTH Aachen University}

\date{February 2023}

\begin{abstract}
The Nielsen-Ninomiya theorem, dubbed `fermion-doubling', poses a problem for the naive discretization of a single (massless) Dirac cone on a two-dimensional surface.
The inevitable  appearance of an additional, unphysical fermionic mode can, for example, be circumvented by introducing an extra dimension to spatially separate  Dirac cones. 
In this work, we propose a geometry-independent protocol based on a
tight-binding model for a three-dimensional topological insulator on a cubic lattice. 
The low-energy theory, below the bulk gap, corresponds to a Dirac cone on its two-dimensional surface which can have an arbitrary geometry.
We introduce a method where only a thin shell of the topological insulator needs to be simulated. 
Depending on the setup, we propose to gap out the states on the undesired surfaces either  by breaking the time-reversal symmetry or by introducing a superconducting pairing.
We show that it is enough to have a thickness of the topological-insulator shell of  three to nine lattice constants. 
This leads to an effectively two-dimensional scaling with minimal and fixed shell thickness. 
We test the idea by comparing the spectrum and probability distribution to analytical results for both a proximitized Dirac mode and a Dirac mode on a sphere which exhibits a nontrivial spin-connection.
The protocol yields a tight-binding model on a cubic lattice simulating Dirac cones on arbitrary surfaces with only a small overhead due to the finite thickness of the shell.
\end{abstract}

\maketitle

\section{\label{sec:intro}Introduction}
A long-standing problem in physics is the efficient simulation of massless Dirac fermions.
Contrary to systems with non-relativistic kinetic energy, the introduction of a lattice and straightforward discretization of space does not lead to a correct low-energy description of the Dirac cone. 
At the core of the problem is the fermion doubling theorem \cite{Nielsen1981} which predicts the inevitability of additional unphysical low-energy modes. 
Mathematically, the doubling can be traced back to the discretization of a first- instead of second-order differential equation.

Over the years, various remedies have been put forward that can be clustered into two general categories according to the dimensionality of the underlying model. The first category contains true two-dimensional models, such as Wilson fermions \cite{Wilson1974,Ginsparg1982} or staggered fermions \cite{Kogut1975,Stacey1982,Pacholski2021}. These are computationally efficient but complicate the description and bandstructure of the system. 
Moreover, a discretization method of the transfer matrix has been developed to solve quantum transport problems in open systems \cite{Beenakker2008,Beenakker2010,Eric2011,Caio2012}. 
While these methods work well for a flat 2D surface, their generalization to curved surfaces  is not at all straightforward \cite{Brower2017}.
One problem is that on a curved space the proper spin-connection has to be taken into account which arises due to the coupling of spin to momentum.

The second category contains higher-dimensional systems in either space \cite{Kaplan1992,Shamir1993,Aoki2022,Aoki2023} or time \cite{Kim2023}. In this approach, the Dirac equation arises as the surface model of a gapped bulk system.  Upon discretization, these systems circumvent the fermion doubling by  separating the Dirac cones. A prominent example are three-dimensional topological insulators where the bandstructure of the surface modes is described by a Dirac cone \cite{Bernevig2006}.  These models allow for a surface of arbitrary geometry. They do so at the expense of less computational efficiency due to the addition of an extra dimension.

The goal of this work is to introduce a method to efficiently simulate Dirac fermions on a two-dimensional surface $S$ of arbitrary geometry. To this end, we follow the second category and embed the system in three-dimensional space.
In particular, we use a topological insulator (TI) \cite{Hasan2010,Zhang2011} whose boundary  coincides with the surface $S$. 
However, instead of filling the entire volume $V$ enclosed by $S$ with the bulk of a TI, we only consider a small shell $\Delta V$ of finite thickness $d$. This procedure creates an additional inner surface $S_a$ with new surface modes. 
To avoid hybridization corrections to the surface physics on $S$, the thickness of the shell must be larger than the decay length of the surface modes $\lambda$ into the bulk. 
The loss of computational efficiency is given by the size of $d$.
The thickness $d$ is  minimized in our approach by locally introducing a term in the Hamiltonian that gaps out the surface mode on $S_a$. 
The selection of this term depends on the details of the system. 
The inclusion of such a term allows for the reduction of the thickness to $d \simeq \lambda$. We show that by correctly tuning the parameters of the TI model, a thickness of 3--9 lattice constants is enough to accurately model the surface states on $S$. Thus, we achieve an effectively two-dimensional scaling. 
This procedure  gives a (local) tight-binding model for which existing packages, like Kwant \cite{Groth2014}, can be employed. 
Thus, both the calculation of the spectrum as well as of transport properties can be delegated to dedicated and optimized packages.

The overview of the paper is as follows. In Sec.~\ref{sec:model}, we introduce the tight-binding model for the TI. We discuss the optimal  choice of parameters and the  low-energy surface model that results. 
We present a protocol to efficiently simulate an isolated Dirac cone based on gapping out the undesired modes on additional surfaces. 
This protocol is intentionally kept general to convey the principles of the method and enable its transfer to  platforms and geometries not studied in this paper. 
We propose two general methods and test them in Sections \ref{sec:proximitized_TI} and \ref{sec:Dirac_sphere}. 
These examples are chosen such that a comparison to analytical results is possible. 
In Sec.~\ref{sec:proximitized_TI} the surface of a TI is proximitized by an $s$-wave superconductor, and a circular surface region is left uncovered. For such a system, Andreev bound states form below the superconducting gap and are localized in the bare surface region. Both the probability density as well as the spectrum of the Andreev states closest in energy to the center of the superconducting gap are computed and found to be in agreement with analytic calculations. This setup can be readily extended to simulate vortex Majorana bound states in this heterostructure \cite{Fu2008,Hasan2010,Ioselevich2012, Akzyanov_2014,Flicker2019,Deng2021,Ziesen2021}. In Sec.~\ref{sec:Dirac_sphere}, a Dirac sphere is simulated, where spinful two-dimensional Dirac fermions are restricted to the curved surface of a two-sphere. The probability density and the spectrum of the finite-size quantized states closest to charge neutrality are simulated and tested by comparing to analytical results. The conclusion and outlook is deferred to Sec.~\ref{sec:conclusion}.

\section{\label{sec:model}The gapped-shell model}

In this section, we present the tight-binding model of the TI that is used and detail the required parameter choices. Furthermore, we describe the inclusion of the boundary effects to obtain the effective two-dimensional model for an arbitrary surface $S$. As a model of a  three-dimensional topological insulator, we take the 3D Bernevig-Hughes-Zhang (BHZ) model \cite{Bernevig2006,Fu2007}
\begin{align}\label{eq:lattice_BHZ}
    H_\mathrm{BHZ} =& M\tau_z- 2B\sum_j [1-\cos(a k_j)]\tau_z \notag \\
   & + A\sum_{j} \sin(ak_j)  \sigma_j\tau_x \, ,
\end{align}
with the wave vector $\bm{k}=(k_x,k_y,k_z)^T$ and  $j= x,y,z$. 
The model can be realized on a cubic lattice with 4 degrees of freedom per unit cell (2 for the spin $\sigma$ and 2 for the orbital $\tau$); in the following, the length scales are measured in multiples of the isotropic lattice constant $a=1$. 
The parameter $A$ is proportional to the linear velocity $v_D$ of the surface modes \footnote{In general, the velocity parameters $A_j$ can be different for each direction. For simplicity, we keep the velocities  isotropic in this work.},  $M$  and $B>0$ are mass terms. 
The system described by the Hamiltonian transitions as $M$ changes sign between a trivial ($M<0$) and topological ($0<M<4B$) phase.
The Hamiltonian is time-reversal symmetric with $[H_\mathrm{BHZ},T]=0$, where $T=i\sigma_y K$ ($K$ denotes complex conjugation). 

\subsection{Optimal parameters and surface Hamiltonian \label{subsec:TI_model}}
Since we are only interested in an accurate description of the low-energy surface modes around $\bm{k}= 0$, we specifically tune the parameters in the tight-binding model to increase the bulk gap and thus decrease the decay length of surface modes, $\lambda$.
 Evaluating the bandstructure of Eq.~\eqref{eq:lattice_BHZ} at the high symmetry points of the Brillouin zone (in the topological phase with $0<M<4B$), we find the optimal bulk gap of $M$ for $B=M/2$ and $A\geq M$. 
The decay length of the surface modes is then given by $\lambda\approx A/M$. To simulate the bulk as efficiently as possible and minimize the thickness $d$ of the 3D model, we want to minimize the decay length. Without sacrificing the bulk gap, the optimal parameters $A=M = 2B$ lead to a decay length of  $\lambda\approx 1$. A thickness of three lattice spacings is sufficient to suppress the wave function by a factor of $e^{-3} \approx 5\,\%$ compared to the  value on the surface $S$.

\begin{figure}[tb]
  \centering
\includegraphics{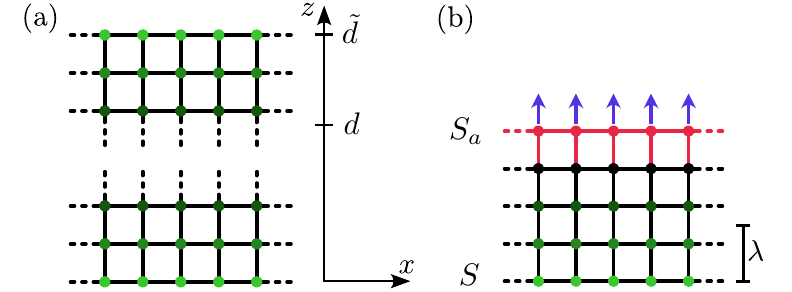}
\caption{Cross section (at $y$ constant) of a cubic lattice that can be used to model the Dirac equation on the flat plane  $S$ ($z=0$). Panel (a): The bulk has a width $\tilde d$. The surface modes decay in $z$-direction on a scale $\lambda$ from the surfaces (green) into the bulk (black) from both surfaces at $z=0,\tilde d$. As  the surface mode at $z=\tilde d$ is unwanted, we need $\tilde d \gg \lambda$ and care has to be taken that it does not influence the result on $S$. A more efficient model is shown in panel (b), where the additional surface $S_a$ at $z=d < \tilde d$ (red) is gapped out, \textit{e.g.}, by an artificial magnetic field (in the direction of the blue arrows). The gap on the upper boundary is improved by increasing the hoppings on the links (red) involving at least one surface site. }
\label{fig:boundary_sketch}
\end{figure}
The only purpose of the BHZ model is to produce the correct Dirac equation on its surface. To fix ideas, we focus on a flat interface in $z$-direction with a topological insulator at $z\geq 0$ and vacuum at $z<0$ [see lower surface in Fig.~\ref{fig:boundary_sketch}(a)]. The corresponding outward surface normal is $\bm{\hat{n}}=(0,0,-1)^T$. To determine the low-energy modes for small $k=|\bm{k}|$, which corresponds to long wavelengths, we expand Eq.~\eqref{eq:lattice_BHZ} to quadratic order in $k$  and analytically continue the bandstructure around $\bm k = 0$ to $k_z = i \kappa$ with $\kappa >0$ in order to find the zero energy modes decaying for $z>0$. Projecting onto the space of surface modes, which consists of the two $+1$ eigenstates of $\sigma_z \tau_y$, yields the effective Hamiltonian
\begin{equation}\label{eq:Hsurface}
H_{\mathrm{sur}} = A\left( k_x \sigma_y - k_y \sigma_x\right) = A (\bm \sigma \times \bm k)\cdot \bm {\hat n} \, .
\end{equation}
This describes a Dirac cone around $\bm{k}=0$ with the Dirac velocity given by the parameter $A$ \footnote{For $k_x,k_y \gtrsim \pi/4$ the lattice simulation deviates from the continuum model due to the replacement $k_j \mapsto \sin(k_j)$ for $j\in \lbrace x,y\rbrace$.
Therefore, the velocity of the surface modes decreases as the bulk gap is approached.
This effect can be accounted for by renormalizing the surface mode velocity for higher energies.} and a spin that is locked to the momentum. The reformulation in the second step of Eq.~\eqref{eq:Hsurface} shows the extension of the Hamiltonian to differently oriented surfaces.
In general, due to the basis selection in Eq.~\eqref{eq:lattice_BHZ} the modes on a surface with an outward-pointing normal vector $\bm{\hat n}$  are the $+ 1$ eigenstates of $-(\bm{\hat n}\cdot \bm \sigma)\tau_y$.
The spin of the surface modes in turn are oriented parallel to the surface. 
Note that the gap-closing at $\bm k =0 $ in Eq.~\eqref{eq:Hsurface} is protected by the time-reversal symmetry.
Adding a symmetry-breaking term $G\, \bm{\hat n}\cdot \bm \sigma$ that corresponds to a magnetic field perpendicular to the surface will open a gap in the spectrum, a fact that will be used in the following.

In summary, using the lattice model of the three-dimensional topological insulator in Eq.~\eqref{eq:lattice_BHZ} and the parameter choice discussed above, Dirac cones around $\bm{k}= 0$ can be simulated with a bulk penetration depth of $\lambda \approx 1$. Figure \ref{fig:boundary_sketch}(a) shows a sketch of a cross section at fixed $y$ through a three-dimensional tight-binding model on a cubic lattice with finite $z\in [0, \tilde d]$ that is extended in the $x$-direction. The two-dimensional Dirac surface modes at $z=0$ are captured by the Hamiltonian in Eq.~\eqref{eq:Hsurface}. However, as there are low-energy modes at all interfaces of the TI with a trivial region (\textit{e.g.}, at finite $z=\tilde d>0$), the bulk size needs to be kept large enough to avoid hybridization between any of the surface modes. This inter-surface hybridization limits the thickness $\tilde d$ of the model orthogonal to the surface and thus the computational efficiency. In the next section,  we circumvent this limitation through intra-surface hybridization by introducing additional terms in the Hamiltonian that gap out the modes on the additional surfaces.

\subsection{Gapping unwanted surface modes\label{subsec:gapping}}

In this section, we present a protocol to determine  an optimized model to simulate  Dirac fermions on specific surfaces. The steps consist of the choice of the gapping term, tuning of the gapping parameters, choice of boundary conditions for the finite system and a generalization of the procedure for an arbitrary geometry. 

To couple the surface modes described by Eq.~\eqref{eq:Hsurface}, we  follow two  methods. The \emph{mode duplication method} (MDM) is inspired by superconductivity. We can double the degrees of freedom by adding an (artificial) duplicate of the surface model to our description. Coupling the unwanted mode on $S_a$ between the system and its duplicate gaps out the surface mode. For the \emph{local coupling method} (LCM), we keep the number of degrees of freedom constant and couple the two modes within a surface by a `magnetic field' that breaks the time-reversal symmetry and opens a gap. In particular, the magnetic field must have a component perpendicular to the surface in order to open a gap. Because of this, the coupling terms have to be position-dependent for the LCM \footnote{Note that the LCM therefore is local in position space, which is different from the Wilson mass that is local in momentum space.}.

The MDM is rather easy to implement. We introduce an extra local degree of freedom with Pauli matrices $\eta_j$ acting on it. The surface states are gapped out by considering the Hamiltonian ($G>0$)
\begin{equation}\label{eq:Hlat_isotrop_gap}
    H'_\mathrm{BHZ} = H_\mathrm{BHZ}\,\eta_z + G \eta_x\, ,
\end{equation}
acting on the enlarged Hilbert space. Thereby, a gap of size $2G$ is opened symmetrically around $\epsilon=0$ in the spectrum of $H_\mathrm{BHZ}$. Physically, this Hamiltonian corresponds to coupling the BHZ system to an  $s$-wave superconductor with pairing-strength $G$. In this way, the time-reversal symmetry of the system is retained. The projection on the surface mode is not affected by the additional degree of freedom. We obtain
\begin{equation}
    H'_\mathrm{sur} = H_\mathrm{sur}\eta_z + G\eta_x\, .
\end{equation}
As a result, the surface states become gapped due to the mass term $G \eta_x$. 
Adding the mass term only on the unwanted surface $S_a$ [red sites in Fig.~\ref{fig:boundary_sketch}(b)], the low-energy model corresponds to a (massless) Dirac equation only on the green surface $S$.
For the black and green sites, we do not couple the two lattices and set $G=0$. On the other hand, for the red sites, we locally choose $G>0$. In this case for energies $|\epsilon|<G$ surface states only exist on $S$ so that no hybridization with surface states on $S_a$ is possible. Therefore, in the aforementioned energy regime, the opposite surface can be ignored and the intended surface physics on $S$ is simulated already accurately with a shell of  thickness $d = 4$ [see Fig.~\ref{fig:boundary_sketch}(b)].

For the LCM, we need to incorporate a local term to \eqref{eq:lattice_BHZ} such that on the unwanted surface $S_a$ a magnetic field perpendicular to the surface is produced which gaps the surface modes.  We thus choose
\begin{equation}\label{eq:Hfull_gap_mag}
    H'_\mathrm{BHZ} = H_\mathrm{BHZ} + G  \,\bm n \cdot \bm \sigma \, ,
\end{equation}
with $\bm{n}=(n_x,n_y,n_z)^T$ the local outward-directed surface normal and $G>0$. 
The implementation is best understood by again studying the sample system with a surface at fixed $z=d$ [cf. Fig.~\ref{fig:boundary_sketch}(b)]. For the artificial surface $S_a$ (red) the surface normal is $(0,0,1)^T$  (blue arrows) leading to a gapping term $G\sigma_z$ and a low-energy mode governed by the last expression in \eqref{eq:Hsurface}. For the projection onto the low-energy sector on this specific surface, we obtain
\begin{equation}\label{eq:Hsur_gap_mag}
    H'_\mathrm{sur} = H_\mathrm{sur} +G\sigma_z.
\end{equation}
The spectrum of Eq.~\eqref{eq:Hsur_gap_mag} has a gap of $2G>0$ for the surface states around $\bm{k}= 0$. Analogous to the procedure of the MDM, we only introduce the gapping term with $G>0$ on the upper surface $S_a$ (red sites) to prevent the surface states on the simulated surface $S$ (green) for $|\epsilon|<G$ from hybridizing. Thereby, we can again limit the shell thickness to $d=4$.

The remaining question is how to optimize the gapping of the modes at $S_a$ without affecting the physics on $S$. Simply increasing $G$ until it reaches the value of the bulk gap $M$ does not work as intended. We observe the surface states to move one layer inwards into a former bulk layer for $G > M/2$. A similar behavior of inward motion for the surface states is observed in the case of strong surface disorder \cite{Schubert2012}. 
To force the states back into the surface layer to get affected by the coupling term, we increase the hopping into and within the artificial boundary region depicted by the red links in Fig.~\ref{fig:boundary_sketch}(b). To achieve this, we introduce additional scaled TI parameters $(M_G, B_G, A_G)$ on the red sites and links, where it is clear from the previous Sec.~\ref{subsec:TI_model} that the ratio of the TI parameters needs to stay fixed to preserve bulk properties. We have found that the choice of $G=100M$ and $M_G=A_G=2B_G=10M$ has yielded optimal results by gapping the   surface states  up to the bulk gap \footnote{The occurrence of artificial low energy states due to the sharp transition of the TI parameters can in general be present. However, we have not found such artifacts when following the procedure presented in the main text.}.
\begin{figure}[tb]
  \centering
\includegraphics{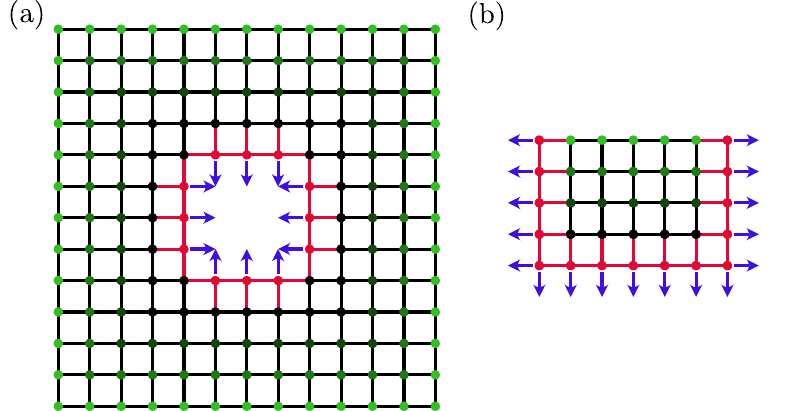}
\caption{Cross section of a tight-binding model for different topologies of the surface that are aligned with the lattice (color scheme  as in Fig.~\ref{fig:boundary_sketch}). Panel (a): The system (green) models a closed surface. In order to simulate an open topology, as in panel (b), the artificial surface (red) is neighboring the system.}
\label{fig:closing_sketch}
\end{figure}
\subsection{Simulating arbitrary geometries}

So far, we have only studied a simple half-plane. Without periodic boundary conditions, the surfaces of the TI have to be closed as it is surrounded by the vacuum which corresponds to the trivial phase. This means that we either simulate a closed (green) surface, cf. Fig.~\ref{fig:closing_sketch}(a) or Sec.~\ref{sec:Dirac_sphere} for an example, or we have to gap out also part of the outer surface to be left with an open surface, cf. Fig.~\ref{fig:closing_sketch}(b) and see Sec.~\ref{sec:proximitized_TI} for an example. 
In the latter case the direct coupling  of the surface modes (green region) to the gapped surface (red) is also increased to the value of the red links in order to impose  hard-wall boundary conditions. 
This allows us to truncate the simulated system after a single gapped surface site. The selection of the closing procedure is determined by the efficiency  for the specific physical platform that is simulated.

\begin{figure}[tb]
  \centering
\includegraphics{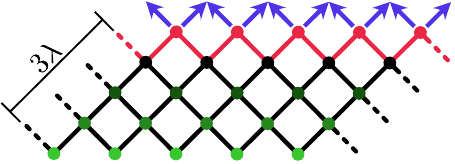}
\caption{Tilted surface that can be used to implement arbitrary geometries. Here, the specific case of a 45$^\circ$ tilting is shown. Multiple neighboring sites are missing at the upper surface. The surface normal is given by the sum of the directions to the missing neighbors.}
\label{fig:geom_sketch}
\end{figure}
The MDM is readily adjusted as the term $G\eta_x$ gaps the surface mode in an arbitrary direction. For the LCM, we need to adjust the direction of the magnetic field according to the orientation of the surface, see Fig.~\ref{fig:geom_sketch}.
We have found that the following simple and local method works  well: at each lattice site, we sum the vectors pointing to  missing neighbors. This yields a vector $\bm n$ that approximates the normal vector at this site.
We then add the onsite potential $G \,\bm n \cdot \bm \sigma$.
Note that, for optimal results, we do not normalize the vector $\bm n$ such that the strength of the effective magnetic field $G \bm n$ depends on the number of missing neighbors.
This method only relies on the (local) knowledge of the boundary points and missing neighbors of the cubic lattice and as such can be easily implemented for arbitrary geometries.

Before demonstrating the described method for two explicit examples, we comment on the main sources of errors. The simulated surface states have an exponentially small probability $\propto e^{-d/\lambda}$ to be found at the artificial surface. This has two potential consequences. First, the surface states obtain a small gap leading to an error in simulated energies $\mathcal{O}\left(e^{-d/\lambda}\right)$. This error has to be seen in relation to the discretization error, where system scales are compared to the lattice scale. As a second consequence, since the LCM locally breaks time-reversal symmetry, the simulated surface modes only approximately preserve time-reversal, with an error again exponentially decaying in $d$. 

In the following, we test our method on two examples. The examples are chosen to show the versatility of the method while still allowing the comparison of the results to analytics.
Whether to choose the MDM or LCM is a matter of convenience. 
The MDM is easier to implement as it does not require knowledge of the normal vector.
However, this comes with the drawback of doubling the degrees of freedom.
As a result, we propose to use MDM when simulating a superconducting system as in Sec.~\ref{sec:proximitized_TI}, where doubling is required anyway.
For normal-conducting setups, as in Sec.~\ref{sec:Dirac_sphere}, we consider LCM to be the method of choice.

\section{\label{sec:proximitized_TI}Proximitized topological insulator}

\begin{figure}[tb]
  \centering
\includegraphics{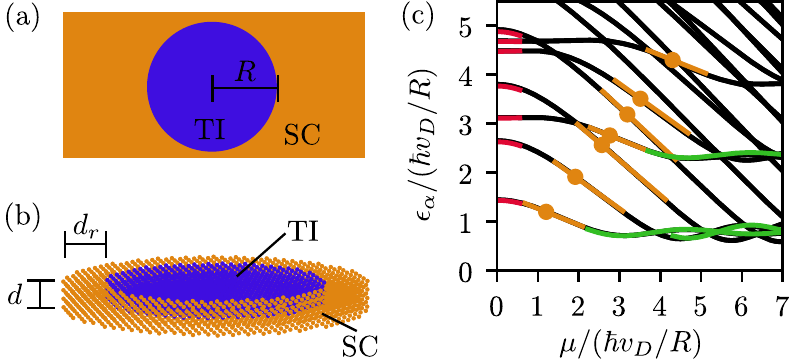}
\caption{ Panel (a): The surface of a TI (blue) is covered by an $s$-wave superconductor (orange) except for a disk of radius $R$. Panel (b): A lattice model that corresponds to the system in (a). A small bulk of thickness $d$ is added to the surface layer. The sides and bottom surface are gapped out by superconducting pairing. The system has a radius of $R+d_r$. Panel (c): Simulation of the low-energy spectrum of the system as a function of the surface chemical potential $\mu$ (black curves). All energies are doubly degenerate due to the time-reversal symmetry. They are in good agreement with analytic predictions for $\epsilon_\alpha\gg \mu$ [red, see \eqref{eq:TI_SC_smallmu}], $\epsilon_\alpha \approx \mu$ [orange, see \eqref{eq:TI_SC_mediummu}] and $\epsilon_\alpha \ll \mu$ [green, see \eqref{eq:TI_SC_largemu}]. The orange dots mark the cross-over points $\epsilon_\alpha=\mu$ between the different regimes. The parameters are $R=24.5$ and $d=d_r=3$ in units of the lattice spacing. } 
\label{fig:SC_TI_sketch}
\end{figure}

In this first application, we simulate a TI whose surface is partially proximitized by a superconductor, see Fig.~\ref{fig:SC_TI_sketch}(a).
Note that we would like to simulate the Dirac surface mode in the disk of radius $R$ that is proximity coupled to the superconductor at its boundary.
The Hamiltonian of the system can be described in Bogoliubov-de Gennes formalism with $H = \tfrac{1}{2} \int d^2r\, \Psi^\dagger(\bm{r}) H_{\mathrm{BdG}} \Psi(\bm{r})$. As a basis, we choose $\Psi(\bm{r})=$ $ [\psi_\uparrow(\bm{r}),\psi_\downarrow(\bm{r}),\psi^\dagger_\downarrow(\bm{r}),-\psi^\dagger_\uparrow(\bm{r})]^T$ with the fermionic field operators $\psi_\sigma$, where $\sigma$ is the spin degree of freedom of the surface modes. 
The matrix Hamiltonian for the proximitized TI surface is \cite{Hasan2010}
\begin{equation}\label{eq:hamiltonian_SC_TI}
    H_{\mathrm{BdG}} = ( v_D \,\bm{p}\cdot \bm{\sigma}-\mu)\eta_z + \Delta(\bm{r})\,\eta_x \, ,
\end{equation}
with $\bm{p}=(p_x,p_y)^T$, the surface velocity $v_D$, the chemical potential $\mu$ and the superconducting pairing profile $\Delta (\bm{r})$.
We simulate the system without a magnetic vortex such that the pairing term is real-valued.
For the gap profile, we take $\Delta(\bm{r}) = \Delta\,\Theta(|\bm{r}|-R)$ with $\Theta(x)$  the Heaviside step-function. 
The Hamiltonian has particle-hole symmetry $\{ H,P\}=0$, with the particle-hole operator $P=\sigma_y\eta_y K$.

The problem lends itself to the MDM. In particular, as we are interested in the modes well below the superconducting gap $\Delta$, we are free to increase $\Delta$ as much as we want (which then approaches a hard-wall boundary condition).
We choose $\Delta = G = 100M$ in order to obtain a decay length $\xi \simeq A/G \ll 1$ of the surface modes into the proximitized region with $|\bm r| \geq R$.
Being only interested in the correct low-energy description, we are free to minimize the bulk of the TI that we simulate.
Figure \ref{fig:SC_TI_sketch}(b) depicts the minimal tight-binding model that captures the correct low-energy physics.
The physical system that is to be simulated is represented by the upper-most layer, where a ring of width $d_r$ with superconducting pairing $\Delta> 0$ traps the surface states in the unproximitized TI region. 
This cross section is continued for four layers in the bulk direction ($d=3$) in order to allow for the surface modes to decay.
The physical system of approximately  $N_\text{ph}=\pi R^2$ lattice points is embedded in a tight-binding model with $N_\text{tot}=\pi (R+d_r)^2 (d+1)$ lattice sites with $d_r = d =3$.
For large $R$, we find $N_\text{ph}/N_\text{tot} \approx (d+1)^{-1}$, \emph{i.e.}, approximately only a fourth  of the lattice sites used in the simulation are `part of the system'. 
Because of this scaling, it is important to keep $d$ as small as possible, which explained the detailed fine-tuning of the parameters in Sec.~\ref{subsec:gapping} in order that $\lambda \approx 1$.

As mentioned above, we are only interested in the surface states well below the superconducting gap. For those states with energy $|\epsilon| \ll \Delta $ the exact value of the superconducting pairing is irrelevant and we may choose $\Delta =G$ throughout the orange region.   
If the physics for $|\epsilon| \lesssim \Delta$ is of interest, a gradient in $\Delta$ can be introduced along the bulk direction and the width of the proximitizing ring has to satisfy $d_r > 3\xi = 3 \hbar v_D /\Delta$ to correctly capture the surface physics. 
A code example for the implementation of the tight-binding model in Kwant, where we chose $d_r=d$ is available at \cite{zenodo}.

The simulation results for the twelve lowest energy states as a function of the surface chemical potential $\mu$ are depicted in Fig.~\ref{fig:SC_TI_sketch}(c). The time-reversal symmetry of Eq.~\eqref{eq:hamiltonian_SC_TI} is reflected in the occurrence of Kramers' pairs, leading to a two-fold degeneracy of the energy levels. Due to the particle-hole symmetry the levels are symmetric around $\epsilon=0$ such that only the positive energies are shown. 

The spectrum $\epsilon_\alpha$ of the Hamiltonian Eq.~\eqref{eq:hamiltonian_SC_TI} is analytically obtained  in Appendix \ref{app:analytic_approx}. 
We find that due to the angular symmetry of the system, each state can be labeled by a tuple $\alpha = (m,n)$ of the angular and the radial quantum number. For $\epsilon_\alpha\gg \mu$ the simulated curves accurately follow the analytic prediction in \eqref{eq:TI_SC_smallmu} to second order in $\mu$. With increasing $\mu$ the energies $\epsilon_\alpha$ tend to decrease and a cross-over regime is reached for $\mu \approx \epsilon_\alpha$. At this point, the chemical potential becomes large enough that the distance to charge neutrality is larger than the quantized energy of the state $\alpha$ leading to electron-like states. 

Around the cross-over points, with $\mu \approx \epsilon_\alpha$, the simulated spectrum closely follows the analytic prediction \eqref{eq:TI_SC_mediummu} of linear decay. The slope of the decay is purely dependent on the angular quantum number  $m$. 
In the regime $\epsilon_\alpha \ll \mu$, the results are only valid for $\mu\lesssim M/2$ as otherwise unwanted effects due to the bulk-modes play a role. The range of chemical potential in Fig.~\ref{fig:SC_TI_sketch}(c) with $R=24.5$ corresponds to $\mu \in [0,0.3M]$. 
In order to increase the range of $\mu$, a larger radius has to be simulated \footnote{The value $R=24.5$ is chosen to show the memory efficiency of the presented approach. This system was run with only 3GB RAM usage on a Windows machine without MUMPS, despite it being a three-dimensional setup.}. 

The analytic result in \eqref{eq:TI_SC_largemu} shows that for $\mu\to\infty$ the energies $\epsilon_\alpha$ go towards a limiting value that is independent of the angular quantum number $m$. Indeed, we observe the clustering of the energies $\epsilon_\alpha R/\hbar v_D$ in Fig.~\ref{fig:SC_TI_sketch}(c) for large $\mu$ around $\pi/4$ (for  $n=0$) and $3\pi/4$ (for $n=1$). 
The first correction in this limit is captured by sinusoidal oscillations (that depend on $m$). 
The simulated oscillations for the lowest three initial states  are in good agreement with the analytical prediction (green). Generally, we found that the discrepancy between simulation and analytic results exponentially converges in $d$ towards the discretization error $\mathcal{O}\left( 1/R\right)$.

\begin{figure}[tb]
  \centering
\includegraphics{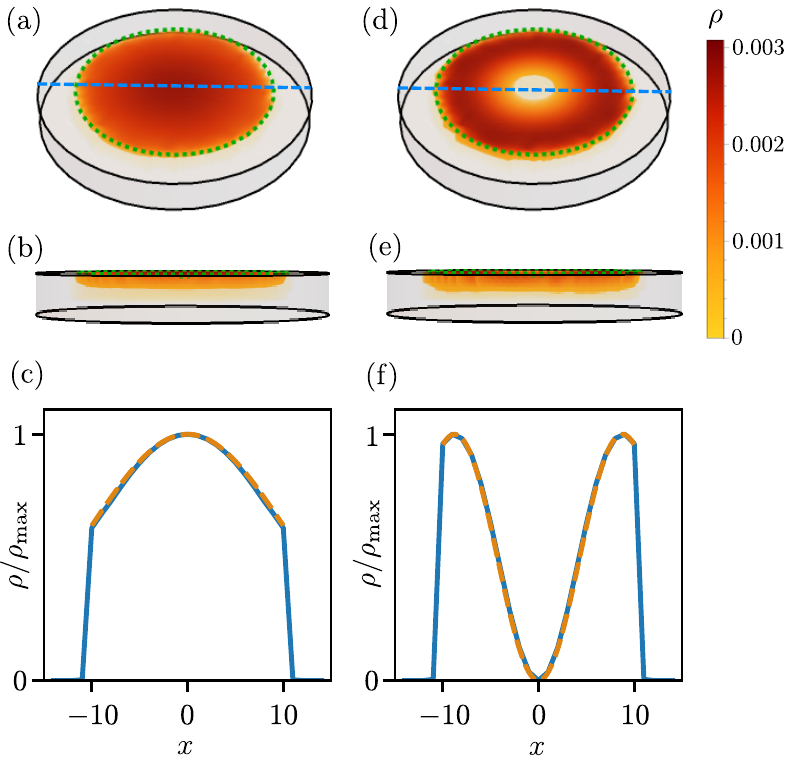}
\caption{Panels (a)--(c): Probability density of the lowest energy mode of the proximitized TI simulation with $R=10.5$ and $d=d_r=3$. Note that the opacity changes along the color bar in panels (a, b) to allow for a three dimensional view. The complete system that is simulated corresponds to the gray cylinder. Panel (a): The lowest mode is trapped in the bare region of the topological insulator surface (green circle) and does not enter the proximitized sites. The side view, panel (b), shows that the mode also decays rapidly into the bulk. The line cut in panel (c) of the probability distribution $\rho$ for $y=0$ (blue), normalized to its maximal value $\rho_\mathrm{max}$, shows excellent agreement with the analytical prediction (orange dashed) of Eq.~\eqref{eq:wf_TI_box}. Panels (d)--(f): same as panels (a)--(c) but for the first excited state. The mode decays more slowly towards the bottom surface and has one node in the radial direction.}
\label{fig:SC_TI_result}
\end{figure}

Besides the spectrum, the simulation also gives access to the wave functions. This allows to test the localization of the low-energy modes to the physical surface region $S$. In Fig.~\ref{fig:SC_TI_result} we show the probability density $\rho=|\psi|^2$ of the ground [(a)--(c)] and first excited state [(d)--(f)] for $\mu=0$. 
The states are confined to the region of unproximitized topological insulator surface (inside the green circle).
We find a good agreement with the expected decay length $\lambda\approx1$ into the bulk direction normal to the surface.
We observe that along the surface of the model the modes decay even faster, such that we could set $d_r=1$ without compromising the accuracy of the lowest modes.
We find that even though the radius of $R=10.5$ is small, the cubic lattice approximates the circular geometry of the analytical model, see App.~\ref{app:analytic_approx}, rather well. 
In particular, in Fig.~\ref{fig:SC_TI_result}(c) and (f), we compare the probability distribution for the ground state and the first-excited state with analytical results, which demonstrates that the simulation protocol based on the MDM works well for this setup.

\section{\label{sec:Dirac_sphere}Dirac sphere}

\begin{figure}[tb]
  \centering
  \includegraphics{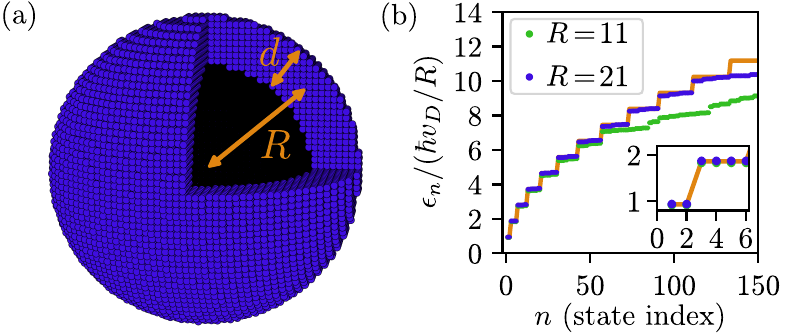}
\caption{Panel (a): Schematics of the tight-binding model for the spherical shell of finite thickness $d$. The artificial surface $S_a$, located at $R-d/2$, is constituted of those blue sites which border the hollow inside (black). A mass profile $M(r)$ smoothly varies from $1$  (topological) at $R-d/2$ to $-1$ (trivial) at $R+d/2$.  The position $R$ of the sign change of the mass defines the location of the Dirac sphere. Panel (b): Results for the eigenenergies $\epsilon_n$ of the finite-size quantized modes on the Dirac sphere. The discrete levels are displayed as a function of the level index $n$ for different radii $R$. For a radius of $R=11$, the first $40$ states agree with the theoretical prediction $\epsilon_j = (\hbar v_D/R)( j + \tfrac{1}{2})$ (orange); note that we had to rescale the vertical axis of the theoretical plot by $7\,\%$, such that the constant heights of the plateaus are in good correspondence. For  $R=21$, already $\approx140$ states agree rather well (blue).}
\label{fig:Dirac_sphere_sketch}
\end{figure}

As a second test of our method, we simulate the spectrum of the Dirac equation on the (two-)sphere. This example is particularly interesting, as a proper modeling of the system entails the simulation of the spin-connection on a curved manifold. 
In spherical coordinates $(r, \theta, \phi)$, the Dirac sphere of radius $R$ is governed by the Hamiltonian \cite{Baer1996,Abrikosov2002,Bagchi2023}
\begin{align}\label{eq:Hamilton_Dirac}
    H_D &= -i \frac{\hbar v_D}{R} \left[ \sigma_x\left( \partial_\theta + \frac{\cot \theta}{2} \right) + \sigma_y \frac{\partial_\phi}{\sin \theta}\right]\, .
\end{align}
The eigenspectrum is given by $\epsilon_j = \pm\frac{\hbar v_D}{R}\left( j + \tfrac{1}{2}\right)$ \cite{Baer1996,Abrikosov2002,Aoki2022}.
Each level has a $2j+1$-fold degeneracy according to $m_j= -j,\dots,j$, the projection of the total angular momentum $j$ along a given direction.
The total angular momentum is a combination of the orbital angular momentum $l=0,1,\dots$ with the spin $\frac12$ of the Dirac particle.  
The degeneracy of each energy eigenspace is even due to the Kramers' degeneracy between the states with $\pm m_j$.
The energy of the system originates from the orbital angular momentum $\bm L$ of the Dirac particle. 
This can be understood as follows:
 $\bm L$ is in an equal superposition of $l^{\pm}=j\pm \tfrac{1}{2}$. 
The average value of $L^2$ is thus given by $\tfrac{1}{2}[l^+(l^+ + 1) + l^-(l^- + 1)] = (j+\tfrac{1}{2})^2$ which coincides with the square of the eigenspectrum measured in units of $\hbar v_D/R$ (for details see \cite{Abrikosov2002}). 

Due to the curved surface, methods that rely on a two-dimensional lattice such as Refs.~\cite{Wilson1974,Kogut1975,Stacey1982} cannot be easily transferred to this setup, as constructing a two-dimensional lattice on the surface of the sphere either has lattice vectors with varying length as in longitude-latitude grids or approximately evenly spaced lattice sites with a strong local variation of the lattice vector direction \cite{Swinbank2006}. 
In contrast, our method can be easily adopted to this setup,  while only producing a small overhead due to the finite thickness $d$ of the spherical shell.

To closely approximate the surface of a sphere, we introduce a mass profile $M(r) = - M \tanh[(r-R)/w]$ smoothly describing the interface between trivial [$M(r)<0$ for $r>R$] and topological region [$M(r)>0$ for $r<R$]. Here, $M>0$ is the mass parameter and $w$ controls the smoothness of the transition. We set $w=1$ throughout this section. The lattice model is sketched in Fig.~\ref{fig:Dirac_sphere_sketch}(a). The outside surface at $R+d/2$ is free of surface modes, because a trivial insulator is in contact with vacuum. In the range $r\in [R-d/2,R+d/2]$ the mass profile $M(r)$ is present, smoothly transitioning between topological and trivial insulator. The sign transition appears at $r=R$ and defines the curved surface $S$ of the two-sphere we intend to simulate. At $R-d/2$ we truncate the lattice model and employ the LCM to remove the low-energy states from the additional interior surface $S_a$.
In Eq.~\eqref{eq:Hfull_gap_mag}, we set $G=B=100M$ with the magnetic field strength only on the inner surface. The outward pointing surface normal $\bm{n}$ is determined according to the rule explained in Sec.~\ref{subsec:gapping}. For the thickness of the ring we set $d=9$ as we have to account both for the decay of the modes towards the trivial $(r>R)$ as well as towards the nontrivial $(r<R)$ side of the transition. 
A minimal code sample for the implementation of this tight-binding model in Kwant is available at \cite{zenodo}.

For radii $R=11$ and $R=21$ the spectra of the simulation for the Dirac sphere are depicted in Fig.~\ref{fig:Dirac_sphere_sketch}(b) \footnote{The model with $R=21$ which resolves the lowest $2\cdot 110$ states, if negative energies are considered, needs 10 GB RAM without MUMPS on a Windows machine.}.
The $140$ lowest and positive energy levels $\epsilon_n$ are plotted against the level index $n$.
Note that the spectrum is symmetric around zero such that negative energies are omitted.
We observe the energy quantization and $2j+1$-fold degeneracy for the states at low energy.
For both $R=11$ and $R=21$ the ground-state space to $j=\tfrac{1}{2}$ is doubly degenerate whereas the state space for the first excited level is four-fold degenerate with $j=\tfrac{3}{2}$.
The plateaus are expected to increase in size for larger $j$, due to a larger degeneracy for higher energy.
We observe that this degeneracy becomes less accurate for larger $j$.
This is due to the fact that for larger $j$, the wave functions have more structure and thus are affected by the corrections due the finite lattice spacing.
As the surface has $N_\text{sur} \approx \pi R^2$ lattice sites, we expect that the number of states that are accurately modeled by the finite system grows as $R^2$.
Indeed, we find that 40 states are approximated well for $R=11$ while already 140 states are captures for $R=21$.
Another way to understand this scaling is to note that only energies with $\epsilon\lesssim M/2$ are well approximated.
Approaching the insulator bulk gap $M$, the surface state velocity gets renormalized, leading to varying step heights.
The number of states with $\epsilon< M/2$ is approximately given by $\tilde n = (M R/2 \hbar v_D)^2$ and thus grows quadratically with $R$.
In particular, we have the estimate  $\tilde n  = 30$ (for $R=11$) and $\tilde n \approx 110$ (for $R=21$) which serves as a good upper bound on the number of states for which our approach works, see Fig.~\ref{fig:Dirac_sphere_sketch}(b). 
We also observe that for $d\geq 9$ the simulation error is dominated by the discretization error of the order $\mathcal{O}(1/R)$ and the time-reversal symmetry breaking of the LCM was exponentially small. 

\begin{figure}[tb]
  \centering
\includegraphics{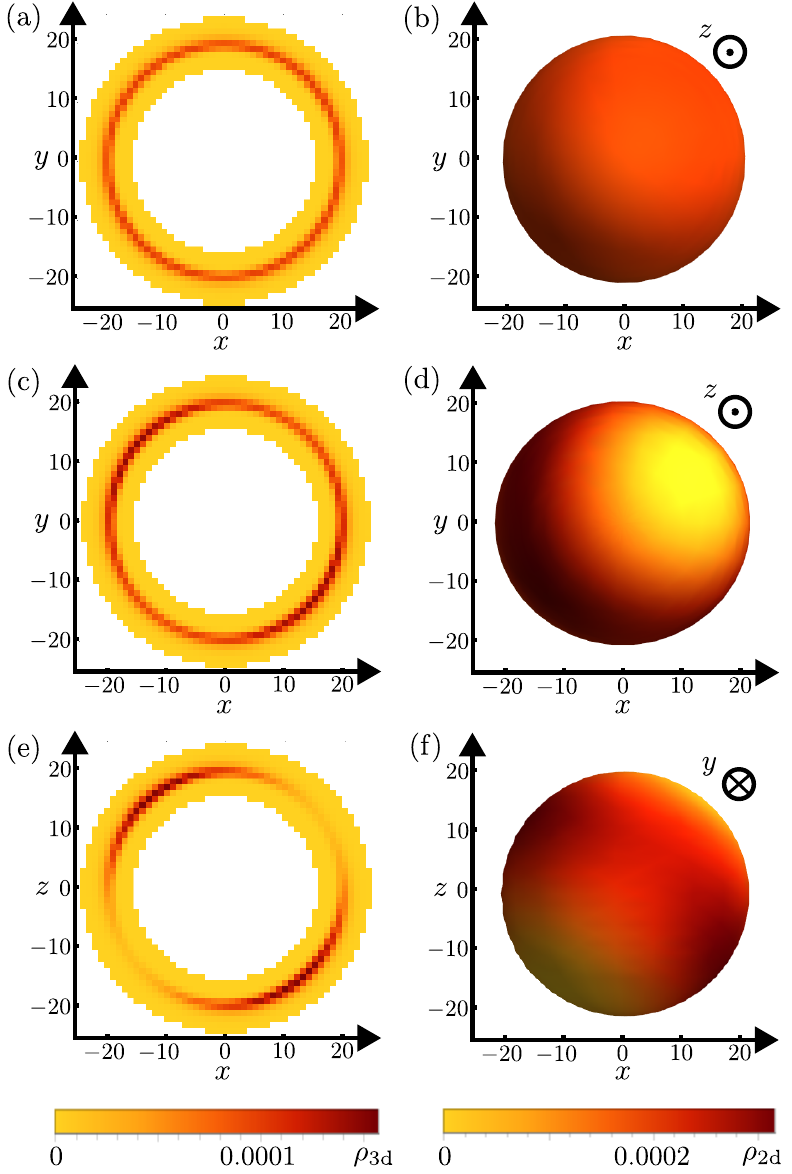}
\caption{Panels (a) and (b) show the probability density of one of the two ground states for $R=20$. Panel (a): Cross-sectional view of the lattice result at $z=0$. The discrete probability density $\rho_\mathrm{3d}$ is approximately isotropic along the circle of radius $R$ and rapidly decays along the surface normal. Panel (b): The marginal surface probability-density $\rho_{\mathrm{2d}}$ from the viewpoint of the positive $z$-direction, after interpolation of the lattice result and integration along the radial direction. The ground state is isotropic on the surface of the sphere.  Panels (c)--(f) are analogous to panels (a) and (b) for one of the first excited states from two different viewpoints: positive $z$-direction in panels (c, d), negative $y$-direction in panels (e, f). In the first excited sector, the probability density has a node along [111], due to the cubic lattice breaking rotational symmetry. This state corresponds to the maximal angular momentum of $\frac{3}{2}$ along [111].}
\label{fig:Dirac_results_sim}
\end{figure}

We visualize the wave functions by plotting the probability density for one state each from the state spaces of the ground and first excited level in Fig.~\ref{fig:Dirac_results_sim}.
The probability density $\rho=|\psi|^2$ of one of the two ground states is depicted in panels (a, b). We observe from the cross-sectional view (a) that for the ring thickness $d=9$ and a mass profile with $w=1$ the state is well localized at $R=20$ within our simulated shell. This enables us to calculate the approximate simulation result for the two-dimensional surface probability-density $\rho_{\mathrm{2d}}(\theta,\phi)$. To achieve this, we interpolate the discrete density $\rho_{\mathrm{3d}}(x_q,y_q,z_q)$ of the three-dimensional eigenstate to obtain a continuous density $\rho_{\mathrm{3d}}(r,\theta,\phi)$. We integrate over the radial direction  to determine the surface probability-density $\rho_{\mathrm{2d}}(\theta,\phi)=\int dr\, r^2 \rho_{\mathrm{3d}}(r,\theta,\phi)$. The result for the ground state is shown in panel (b). It can be seen that the density is spherically symmetric, as expected from the $j=\tfrac{1}{2}$ sector \cite{Abrikosov2002}. For the first excited state (c--f) we find the probability distribution to have a node along [111]. This agrees with theoretical predictions for the $j=\tfrac{3}{2}$ sector \cite{Abrikosov2002}. The quantization axis [111] is fixed by the lattice which breaks the spherical symmetry of the system.

This concludes the discussion of the simulation for the Dirac sphere, where we have proven that with the shell model and local gapping terms the spectrum and probability density of the states at low energy are efficiently computed.

\section{\label{sec:conclusion}Conclusion}
We have introduced a lattice model to simulate an isolated Dirac cone in two dimensions.
The method gives a conventional tight-binding model that can be easily used with existing packages such as Kwant.
The method is efficient as it scales like a two-dimensional problem. Furthermore, it is independent of geometry, allowing the simulation of the spin-connection on a curved manifold.
At the core of it are (efficient)  gapping mechanisms for additional surfaces based on either doubling of the degrees of freedom (\textit{e.g.}, superconductivity) or local symmetry breaking terms (\textit{e.g.}, magnetic fields).
We have tested the method for two setups: a proximitized disk and the Dirac sphere.
For both systems, we showed good agreement for the spectra and probability distribution of the low-energy states close to charge neutrality with analytical predictions.
This gives confidence that, in future work, the method can be transferred to systems where no analytical solutions are known.
As an example, the superconductor-topological insulator heterostructure discussed in Sec.~\ref{sec:proximitized_TI} can be extended to study platforms for Majorana qubits, where a superconducting flux quantum is threaded orthogonal to the surface through the unproximitized region.
The Dirac sphere in turn is extendable to study free fermions on arbitrary two-dimensional geometries.

\section{\label{sec:ack}Acknowledgements}
We acknowledge funding by the Deutsche Forschungsgemeinschaft (DFG, German Research Foundation) under Germany's Excellence Strategy -- Cluster of Excellence Matter and Light for Quantum Computing (ML4Q) EXC 2004/1 -- 390534769 (A.Z. and F.H.) and under Germany’s Excellence Strategy
through the W\"urzburg-Dresden Cluster of Excellence on
Complexity and Topology in Quantum Matter -- ct.qmat
(EXC 2147, project-id 390858490) (C.F.). 

\appendix

\section{Analytical calculation of the proximity coupled disk\label{app:analytic_approx}}
In this section, we present the calculation
of the analytic results for the spectra and wave functions of the Hamiltonian Eq.~\eqref{eq:hamiltonian_SC_TI}. We introduce polar coordinates $(r,\phi)$ relative to the center of the bare TI region and make use of the angular symmetry of the system. For the unproximitized region $r<R$ the eigenvalue equation is solved at energies $\epsilon_\alpha$ with quantum numbers $\alpha=(m,n)$ by the ansatz
\begin{equation}\label{eq:wf_TI_box}
	\psi_\alpha(r, \phi) = \mathcal{N}_\alpha e^{i m \phi} \begin{pmatrix}
		i \,e^{-i\phi/2} J_{m-\frac{1}{2}}[r (\epsilon_\alpha + \mu)/\hbar v_D ] \\
		- \, e^{i\phi/2}  J_{m+\frac{1}{2}}[r (\epsilon_\alpha + \mu)/\hbar v_D ] \\
		i c_\alpha\, e^{-i\phi/2}  J_{m-\frac{1}{2}}[r (\epsilon_\alpha - \mu)/\hbar v_D ] \\
		 c_\alpha\,  e^{i\phi/2} J_{m+\frac{1}{2}}[r (\epsilon_\alpha - \mu)/\hbar v_D ]
	\end{pmatrix} .
\end{equation}
It is obtained analogously to the procedures detailed in \cite{Akzyanov_2014,Deng2021,Ziesen2021} for the case of no flux quantum.
Here, $\mathcal{N}_\alpha $ is a normalization factor, $m$ [$n$] are the half-integer [integer] angular [radial] quantum numbers and $J$ are Bessel functions of the first kind. The values of $c_\alpha$ and $\epsilon_\alpha$ are fixed by continuation of the wave function into the proximitized region $r\geq R$. Since we are evaluating the low-energy modes of the system, we can send $\Delta\to\infty$ and turn the effect of the superconducting region into a boundary condition at $r=R$ that captures the full Andreev reflection. For convenience, we redefine the energies to be measured in units of $\hbar v_D/R$, removing extra constants. The boundary condition then leads to the equations
\begin{multline}\label{eq:energy_TI_SC_implicit}
    J_{m-\frac{1}{2}}(\epsilon_\alpha - \mu)J_{m-\frac{1}{2}} (\epsilon_\alpha + \mu)= \\
    J_{m+\frac{1}{2}}(\epsilon_\alpha - \mu)J_{m+\frac{1}{2}} (\epsilon_\alpha + \mu)
\end{multline}
and 
\begin{equation}
    c_\alpha = \frac{ J_{m+\frac{1}{2}}(\epsilon_\alpha + \mu)}{J_{m-\frac{1}{2}} (\epsilon_\alpha - \mu)} ,
\end{equation}
fixing the energies and particle-hole reflection coefficients, respectively. To test the results of the simulation, we solve \eqref{eq:energy_TI_SC_implicit} approximately for three parameter regimes.
\subsection{Small chemical potential}
In the limit $\epsilon_\alpha\gg \mu$ we  expand the Bessel function of index $\beta$ for small arguments up to quadratic order.
The zeroth order solution $\epsilon^{(0)}_{\alpha}$ follows straightforwardly from Eq.~\eqref{eq:energy_TI_SC_implicit} for $\mu =0$, where the implicit expression simplifies to locating zeros of Bessel functions and their derivative. Expanding Eq.~\eqref{eq:energy_TI_SC_implicit} to second order leads to
\begin{equation}\label{eq:TI_SC_smallmu}
    \epsilon_\alpha = \epsilon^{(0)}_{\alpha}+ f_\alpha\Bigl(\epsilon^{(0)}_\alpha\Bigr) \mu^2,
\end{equation}
with the prefactor
\begin{widetext}
\begin{equation}
	f_\alpha\left(x\right) =  \frac{1}{2} \frac{J_{m+\frac{1}{2}}\left(x \right)J_{m+\frac{1}{2}}''\left( x \right) - \bigl[ J_{m+\frac{1}{2}}'\left( x \right) \bigr]^2 - J_{m-\frac{1}{2}}\left(x \right)J_{m-\frac{1}{2}}''\left( x \right) + \bigl[ J_{m-\frac{1}{2}}'\left( x \right) \bigr]^2}{J_{m-\frac{1}{2}}\left(x \right)J_{m-\frac{1}{2}}'\left( x \right) - J_{m+\frac{1}{2}}\left(x \right)J_{m+\frac{1}{2}}'\left( x \right)} \, .
\end{equation}
\end{widetext}
\subsection{Intermediate chemical potential}
For values $\epsilon_\alpha \approx \mu$, we can  expand the Bessel functions with the argument $\epsilon_\alpha+\mu$ around $2\mu$. For the argument $\epsilon_\alpha-\mu$ we make use of the approximation
\begin{equation}\label{eq:Bessel_small_arg}
	J_{\beta}\left( z \right) \approx \frac{1}{\Gamma (\beta +1)} \left( \frac{z}{2} \right)^{\beta} \, ,
\end{equation}
which is valid for $|z| \ll \sqrt{\beta+1}$ \cite{NIST:DLMF}. The next finite order in the expansion increases by $z^2$, making it irrelevant for the linearization considered in this subsection. Inserting Eq.~\eqref{eq:Bessel_small_arg} into Eq.~\eqref{eq:energy_TI_SC_implicit} and evaluating the remaining Bessel functions in lowest order at $2 \mu^*$ yields for each state characterized by $\alpha$ the cross-over point with $\epsilon_\alpha(\mu^*) = \mu^*$ as zero of a respective Bessel function. Expanding to linear order around $2 \mu^*$ in the Bessel functions with argument $\epsilon_\alpha + \mu$ we find a linear decay
\begin{equation}\label{eq:TI_SC_mediummu}
	\epsilon_\alpha = -\frac{m}{m+1} \left( \mu- \mu^* \right) + \mu^* ,
\end{equation}
the slope of which depends on the angular quantum number $m$. An implicit dependence on the radial quantum number is given by $\mu^*$.

\subsection{Large chemical potential}
In the case of large chemical potential $\epsilon_\alpha \ll \mu$, the Dirac physics connected to the Dirac cone is not very important. To obtain analytical results in this regime,  we expand the Bessel functions for large arguments $(|z| \gg |\beta^2 - \frac{1}{4}|)$  \cite{NIST:DLMF} with
\begin{multline}\label{eq:osz_bessel}
	J_{\beta}( z ) \approx \sqrt{\frac{2}{\pi z}}\Bigl[ \cos \left(z - \frac{\pi}{2} \beta - \frac{\pi}{4} \right)  \\
- \frac{4\beta^2 -1}{8z}\sin\left(z - \frac{\pi}{2} \beta - \frac{\pi}{4} \right) \Bigr]\, .
\end{multline}
To lowest order, the approximation to \eqref{eq:energy_TI_SC_implicit} yields
\begin{equation}
	\epsilon^{(\infty)}_\alpha = \frac{\pi}{2} \left( n +\frac{1}{2}\right) \, , 
\end{equation} 
where $n \in \mathbb{Z}_0$. Thus, in the limit $\mu\to\infty$ the spectrum becomes independent of the angular quantum number $m$. 

The angular momentum only enters the next order correction 
\begin{equation}\label{eq:TI_SC_largemu}
	 \epsilon_\alpha =  \epsilon^{(\infty)}_\alpha+(-1)^n \frac{m}{2\mu} \sin \left( 2 \mu - \pi m\right) ,
\end{equation}
that we obtain by including the first two oscillatory terms from Eq.~\eqref{eq:osz_bessel}.
Thus, the angular quantum number enters as a phase shift in the oscillatory correction.
In Sec.~\ref{sec:proximitized_TI} of the main text, the analytic expressions in Eqs.~\eqref{eq:TI_SC_smallmu}, \eqref{eq:TI_SC_mediummu} and \eqref{eq:TI_SC_largemu} are compared to the simulation results.

\bibliography{refs}

\end{document}